\documentclass[preprint,nofootinbib]{revtex4}
\usepackage{graphicx}
\usepackage{amsmath,amssymb}
\usepackage[colorlinks,citecolor=blue,linkcolor=blue,urlcolor=blue]{hyperref}
\usepackage[babel]{csquotes}
\usepackage{mathrsfs}
\usepackage{enumerate}
\def\be{\begin{equation}}
\def\ee{\end{equation}}

\begin{document}
\title{Quantum Gravity Effects around Sagittarius A*}
\author{Hal~M.~Haggard}
\email{haggard@bard.edu}
\affiliation{Physics Program, Bard College, 30 Campus Rd, Annandale-on-Hudson, NY 12504, USA.}
\author{Carlo Rovelli}
\email{rovelli@cpt.univ-mrs.fr}
\affiliation{CPT, Aix-Marseille Universit\'e, Universit\'e de Toulon, CNRS,
Case 907, F-13288 Marseille, France \\
{\small
and Samy Maroun Research Center for Time, Space and the Quantum.}
 }
%\date{}
\begin{abstract}
\noindent Recent VLBI observations have resolved Sagittarius A* at horizon scales. The Event Horizon Telescope is expected to provide increasingly good images of the region around the  Schwarzschild radius $r_S$ of Sgr A* soon.  A number of authors have recently pointed out the possibility that non-perturbative quantum gravitational phenomena could affect the space surrounding a black hole. Here we point out that the existence of a region around $\frac76 r_S$  where these effects should be maximal.

\vspace{1in}

\begin{center}
\textbf{Essay written for the
Gravity Research Foundation 2016 Awards for Essays on Gravitation\\
Submitted on March 31st, 2016 }
\end{center}

\vspace{1.2in}

%{\small Corresponding authors: Hal M. Haggard}

\end{abstract}

\maketitle

\pagebreak

The detection of the event-horizon-scale structure of Sagittarius A* \cite{Doeleman2008} has opened  an exciting new window of observation on the  universe.  The Event Horizon Telescope is planned to reach an impressive angular resolution of 10 $\mu as$ in the $mm$ and sub-$mm$ wavelengths \cite{Fish2013}.  This should be sufficient for observing detailed features of the near-horizon physics of the Sagittarius A* black hole, such as the shadow and the photon ring predicted by the Kerr geometry \cite{Broderick2014}. 

On the theoretical side, a number of authors, using quite diverse arguments, have recently pointed out that non perturbative quantum gravity effects could leak outside the horizon, a phenomenon not captured by conventional local quantum field theory on a classical geometry.  Examples are the quantum description of black holes as Bose-Einstein condensates \cite{Mazur:2004,Dvali:2015}, fuzzbals \cite{Mathur:2005}, Planck star tunnelling \cite{Rovelli2014,Haggard2014, Barrau:2014a, Barrau:2014b, Barrau:2015, Barrau:2016, Chirstodoulou:2016}, and Giddings's metric fluctuations \cite{Giddings:2014}.  Also, one possible interpretation of the firewall theorem \cite{Almheiri:2012rt}, which has recently raised a lively discussion in the theoretical world, is as an indication that ``something strange" should have to happen at the horizon of a macroscopic black hole, possibly violating the hypotheses of the theorem, which include the validity of local quantum field theory on a given background geometry. 

Giddings, in particular, has observed that the possibility that such quantum gravity effects could lead to observable effects cannot be completely excluded, and listed a number of such possibilities in \cite{Giddings:2014}. For instance, quantum gravitational fluctuations of the metric could disrupt the accretion flow in the near-horizon atmosphere region, distort or suppress the photon ring  \cite{Johannsen2010}, or distort the edge of the shadow \cite{Falcke2000}. For Sagittarius A* the photon ring has a size of 53 $\mu as$ and one might even consider the possibility of relating quantum gravity effects to the preliminary apparent discrepancy between the presently observed size and the expected photon ring size of the hole \cite{Krichbaum2013}.  

Here we point out that there is an argument indicating  a specific region outside the horizon where one might expect quantum gravity phenomena to appear first. This observation is based on an estimate---first appearing in \cite{Haggard2014}---of the spacial dependence of a parameter $q$ measuring the reliability of the classical approximation. 

The usual argument for discarding quantum gravitational effects outside the horizon of a macroscopic black hole depends on the hypothesis that the scale of quantum gravitational effects in a region be determined by the ratio $x=l_P/l_R$ between the Planck length $l_P=\sqrt{\hbar G/c^3}\sim10^{-33} cm$ and the curvature length scale $l_R$ in that region. The second can be estimated for instance using the Kretschmann scalar ${\cal R}^2:=R_{abcd}R^{abcd}=l_R^{-4}$, where $R_{abcd}$ is the Riemann tensor, and for a black hole of mass $M$ this is of the order ${\cal R}\sim GM/r^3$.  In the region immediately surrounding the horizon $l_R\sim GM$. For a stellar mass black hole, $GM$ is of the order of $10^5$ cm and we have an extremely small ratio $x\sim 10^{-38}$, which appears to indicate that quantum gravitational effects are completely negligible. 

The problem with this argument is that it disregards cumulative effects: a small cause can pile up over time and give large effects. In a long time $t$, a small acceleration $a$ produces a large final velocity $v$ that can drive the system by an amount $\delta x\sim \frac12 a t^2$ from where it would have been without the small acceleration.  The order of magnitude of an effective correction of the Einstein equations taking into account quantum gravity is reasonably expected to be proportional to the Planck area, and therefore to $l_P^2$. This is like having an additional driving term, proportional to $l_P^2$, in the dynamics. On dimensional grounds, the vacuum Einstein equations ${\cal R}icci=0$ must therefore be corrected to a form like 
\be
   {\cal R}icci + l_p^2  {\cal R}^2 =0. 
\ee
When small, the quantum term behaves like a linear perturbation. The force, of quantum origin, that drives the field away from the classical solution is $\sim  l_p^2  {\cal R}^2$. Integrating this in proper time $\tau$ can give a cumulative effect of the order $l_p^2  {\cal R}^2 \tau^2$, as for the particle example.  Therefore we remain in the classical region only as long as 
\be
           q  = l_P  \ {\cal R} \ \tau
\ee
is small.  More generally, a relevant parameter for classicality should be, on dimensional grounds,
\be
           q  = l^{2-{b}}_P  \ {\cal R} \ \tau^{b},
\ee
which reduces to the above when $b=1$.  The key point is that $q$ may become of order one close enough to the horizon, and after a sufficiently long time. In other words, there is no reason to trust the classical theory outside the horizon for arbitrarily long times. 

Now, consider a stationary observer at a distance $R$ from the horizon.  If $R$ is large, the smallness of the curvature dominates. Therefore to have quantum effects we must approach the horizon.  However, proper time slows down to zero approaching the horizon (with respect to asymptotic time).  Therefore (in asymptotic time), there must be an intermediate distance from the horizon where $q$ is maximised. Let us determine it.  The proper time $\tau$ of a stationary observer is related to the Schwarzschild time $t$ by the red shift factor
\be
\tau=\sqrt{1-\frac{2GM}{c^2r}}\ t.
\ee
Therefore $q$ depends on the radius as 
\be
q(r)  = l^{2-{b}}_P  \ \frac{m}{r^3} \ \left(1-\frac{2GM}{r}\right)^\frac b2 t^b.
\ee
The maximum in $r$ of this function occurs when, with $r_S=2Gm/c^2$ the Schwarzschild radius,
\be
r=r_S\left(1+\frac b6\right),
\ee
which is a finite distance, but not much, beyond the Schwarzschild radius. This is where quantum effects can first appear.  The most plausible hypothesis is $b=1$, which gives the plausible region for quantum effects to appear to be around $r\sim \frac76 r_S$. The quantum effects appear right where they most reasonably should appear: at a macroscopic distance from the Schwarzschild radius, but close to it, and possibly within the reach of observations. 

The argument given does not \emph{prove} that quantum effects exist in that region.  Furthermore, $q$ remains small for a time which is short compared to the Hawking evaporation time, but is still long on astrophysical scales \cite{Haggard2014}. Thus, the argument should only be taken as a possible  indication of a likely region for quantum effects to first appear.  But, given the current sparse understanding of quantum gravity, the fact that near horizon physics is still observationally largely unexplored, and the pervasiveness of quantum mechanics, it may be good to keep this region in mind. 

\vfil

\bibliographystyle{/Users/carlorovelli/Documents/utcaps}
\bibliography{/Users/carlorovelli/Documents/library}
\end{document}